\documentclass{article}
\usepackage[utf8]{inputenc}
\usepackage{amsmath}
\usepackage{amsfonts}
\usepackage{amssymb}
\usepackage{graphicx}
\usepackage{xcolor}
\usepackage{braket}
\usepackage{makecell}
\usepackage{tikz}
\usepackage{wrapfig}
\usepackage{float}
\usepackage{rotating}
\usepackage{lscape}
\usepackage{hyperref}

\newcommand{\rl}[1]{\left(#1\right)}
\newcommand{\lr}[1]{\left(#1\right)}
\newcommand{\no}{\nonumber}
\allowdisplaybreaks

\begin{document}
	
\baselineskip 24pt

\begin{center}
	
	{\LARGE \textbf{QUANTUM ENTANGLEMENT IN ONE-DIMENSIONAL ANYONS }\par}
	
\end{center}

\vskip .5cm
\medskip

\vspace*{4.0ex}

\baselineskip=18pt

\centerline{\large\rm  H S Mani, Ramadas N, V V Sreedhar}

\vspace*{4.0ex}

\centerline{\it Chennai Mathematical Institute,  SIPCOT IT Park, Siruseri, Chennai, 603103 India} 

\vspace*{1.0ex}

\vspace*{1.0ex}
\centerline{\small   \href{mailto:hsmani@cmi.ac.in}{ \texttt{hsmani@cmi.ac.in}}, \href{mailto:ramadasn@cmi.ac.in}{\texttt{ramadasn@cmi.ac.in}}, \href{mailto:sreedhar@cmi.ac.in}{\texttt{sreedhar@cmi.ac.in}} }

\vspace*{5.0ex}

\begin{abstract}
Anyons in one spatial dimension can be defined by correctly identifying the configuration space of indistinguishable particles and imposing Robin boundary conditions. This allows an interpolation between the bosonic and fermionic limits. In this paper, 
we study the quantum entanglement between two one-dimensional anyons on a real line as a function of their statistics.
\end{abstract}
\section{Introduction}
It is well-known that, in quantum mechanics, the indistinguishability of
particles forces the multiparticle wave functions to be either symmetric (bosonic) 
or antisymmetric (fermionic) under the exchange of any pair of particles. In the 
last few decades it has emerged that in low dimensions it is possible to have more general quantum statistics. The 
classical roots for this can be traced to the non-trivial 
topology of the associated configuration space.  Particles 
which obey these generalised statistics are called anyons, and they interpolate between bosons and fermions. Interestingly, 
these particles appear as collective excitations in fractional quantum Hall systems. In view of this, the quantum mechanical and 
thermodynamic properties of anyons have been extensively studied \cite{wilczek1990fractional} \cite{khare2005fractional}. 

The interest in anyons has been revived recently because of their potential 
application in topological quantum computation \cite{kitaev2006anyons}. 
In topological quantum computation, instead of using qubits one uses 
anyons to store information in their non-trivial wave functions. Since these 
are topologically protected, it is hoped that a topological quantum computer 
leads to fault-tolerant and decoherence-free computation \cite{preskilltqc} \cite{sarma2006topological}. 

However, a completely robust, fault-tolerant physical system is not desirable    
because it does not allow us to store any information, let alone manipulate or
extract it. In view of this, it is important to allow the system to interact with the 
apparatus (environment) in a controlled manner. 

This motivates us to revisit the old problems of anyon quantum mechanics, and study them in the framework of open quantum systems. In particular, we are interested in knowing how the entanglement between two anyons depends on the statistics 
parameter when one of them is considered to be the system, and the other, the environment. 

There are two complexities associated with this problem. First, it is well-known 
that for indistinguishable particles, the standard methods used to quantify the entanglement, like finding the Schmidt rank, taking a partial trace, and finding the von Neumann entropy fail to work.  The main reason 
for this is the non-factorizability of the multi-particle Hilbert space of indistinguishable particles. Various approaches has been proposed to circumvent this problem \cite{balachandran2013algebraic} \cite{ghirardi2005identical} \cite{schliemann2001quantum} \cite{pavskauskas2001quantum}
\cite{franco2016quantum}\cite{wiseman2003entanglement}\cite{killoran2014extracting}\cite{franco2018indistinguishability}\cite{benatti2014entanglement}. Second, these approaches mostly restrict their 
attention to bosons and fermions.

Returning to our problem, we find it useful to follow the information theoretic
approach to quantum entanglement developed by Lo Franco and Compagno \cite{franco2016quantum}. In their work they show how it is possible to define 
the reduced density matrix in a system of  indistinguishable  particles by defining 
an inner product between states belonging to Hilbert spaces with different dimensionalities.  It is straightforward to recast this method in the language of 
second quantization \cite{compagno2018dealing} \cite{lourencco2019entanglement}, which is especially 
suited for our purposes.  Within this framework, we show how the results can 
be generalised to anyons by the simple prescription of using the  anyonic algebra 
for the creation and annihilation operators instead of the bosonic and fermionic 
algebras which are recovered as special cases. \\

The rest of the paper is organized as follows.  

In section 2, we review the information theoretic approach developed by Lo Franco and Compagno, with special emphasis on its reformulation in the language of second quantization. 

In section 3, we review the model of indistinguishable particles on a real line, 
 first studied by Leinaas and Myrheim \cite{leinaas1977theory}. In this model
they first construct the classical configuration space by identifying different configurations which can be obtained by permutations of particle positions, 
and then quantize the system to obtain a wave function that interpolates between 
the bosonic and fermionic limits through a statistics parameter $\eta$ coming 
from the Robin boundary conditions. A second quantization of this model \cite{second} gives rise to an $\eta$-dependent algebra for the creation and annihilation operators, which reduces to the usual  bosonic and fermionic algebras as limiting cases.

In section 4, we use the above results to compute the reduced density matrix and 
the von Neumann entropy of a system of two anyons on a line. 

In section 5 we conclude by giving a summary and an outlook. 

\section{Information Theoretic Approach to Indistinguishable Particles}\label{ita}
In the usual approach, a state of a system of indistinguishable  particles is obtained by first quantizing the system as if the particles were distinguishable, by labelling them. 
We then apply the symmetrization postulate on the product wave functions to get 
bosonic and fermionic states \cite{messiah}. 

It is instructive to restate this in the language of transition amplitudes. For 
example, a two-particle state is simply written as $\ket{\psi,\phi}$, where 
$\psi$ and $\phi$ represent single particle states. For 
indistinguishable particles, this two-particle state should 
be thought of as a holistic entity;  it is not possible 
to say which particle is in which single particle state. Since the particles are not labelled, it is evident that the symmetrization postulate is not invoked. Quantum statistics enters through the definition of the 
inner product of these states. 

For distinguishable particles, an initial state $\ket {\phi,\psi}$ can 
only evolve into the final state, say, $\ket{\varphi, \zeta}$ for which we compute the amplitude. 
But when the particles are indistinguishable, both the final states $\ket{\varphi, \zeta}$ and 
$\ket{ \zeta,\varphi}$ contribute to the amplitude. For the case of bosons and fermions, the 
simple recipe of introducing the right sign to account for the  exchange takes 
care of this complication. 

This ad hoc procedure does not easily generalise to anyons. It is therefore desirable to have a more fundamental approach to the problem where the indistinguishability of the particles is maintained through out. This is the idea behind the  information theoretic approach developed in \cite{franco2016quantum}.  

If $\ket{\varphi,\zeta}$ 
and $\ket{\phi,\psi}$ denote two two-particle states, their inner product is,
\begin{equation}
\braket{\varphi,\zeta|\phi,\psi} = \braket{\varphi| \phi}\braket{\zeta|\psi} + \eta \braket{\varphi| \psi}\braket{\zeta|\phi}. 
\end{equation}
where $\eta =1$ for bosons and $\eta = -1$ for fermions. 

The inner product between states belonging to Hilbert spaces of different dimensionality can also
be defined. If we consider an unnormalized two-particle state, $\ket{\Phi} =  \ket{\varphi_1,\varphi_2}$, 
the inner product with a single-particle state $\ket{\psi}$ is 
\begin{equation}
\bra{\psi} \cdot \ket{\varphi_1,\varphi_2} \equiv \braket{\psi | \varphi_1,\varphi_2} =   \braket{\psi|\varphi_1} \ket{\varphi_2}+\eta \braket{\psi|\varphi_2} \ket{\varphi_1} 
\end{equation}
This is a projective measurement on a single particle, where the unnormalized two-particle state is projected on to $\ket{\psi}$. In a similar manner, the inner product between an $N$-particle state and a single-particle state is also defined. This definition of inner product between states belonging to Hilbert spaces with different dimensions can be used to define the reduced density matrix as shown below.

Let $ \ket{\Phi} $ be a normalized $ N $-particle state. To perform the partial trace we choose a basis $\lbrace \ket{\psi_k} \rbrace$ for the single-particle Hilbert space. The normalized pure state after projecting on to a state $\ket{\psi_k}$ is 
\begin{equation}\label{projsingle}
\ket	{\phi_k} = \frac{\braket{\psi_k|\varphi_1, \varphi_2}}{\sqrt{\braket{ \Pi_k^{(1)}}_\Phi }}
\end{equation}
where $ \Pi_k^{(1)}= \ket{\psi_k}\bra{\psi_k}$. 

Define a one-particle identity operator as $\mathbb{I}^{(1)} = \sum_k \Pi_k^{(1)}$. Then the probability of finding a single particle in  the state $\ket{\psi_k}$ is 
\begin{equation}
p_k = \frac{\braket{ \Pi_k^{(1)}}_\Phi }{\braket{ \mathbb{I}^{(1)}}_\Phi }
\end{equation}
With the knowledge of $\ket {\phi_k}$ and the corresponding probabilities $p_k$, the reduced density matrix is defined as follows
\begin{align}\label{limitk}
\rho^{(1)} = \text{Tr}^{(1)} \ket{\Phi}\bra{\Phi} = \sum_k p_k \ket{\phi_k} \bra{\phi_k}
\end{align}
After obtaining the reduced density matrix, the von Neumann entropy can be calculated as usual,
\begin{align*}
S(\rho^{(1)}) = - \text{Tr}\rl{\rho^{(1)} \log \rho^{(1)}} = -\sum_i \lambda_i \log \lambda_i
\end{align*}
where $\lambda_i$ is an eigenvalue of the reduced density matrix.

\subsection*{Second quantization formalism}
We can recast the above idea in the language of second quantization. If $\ket{\Phi}$ is an $N$-particle state, its inner product with a single-particle state $\ket{\psi_k}$ is \cite{compagno2018dealing}
\begin{align*}
a_{\psi_k} \ket{\Phi} \equiv \bra{\psi_k} \cdot \ket{\Phi}
\end{align*}
Note that since $a_{\psi_k}$ is an annihilation operator, the left hand side of the above equation represents
an $(N-1)$-particle state which, by definition, is the inner product on the right hand side.  As mentioned earlier,
this simple expedient allows us to go beyond bosons and fermions by suitably generalising the operator algebra.
We present this in the next section.
 
 We conclude this section by noting that the expression for the reduced density matrix in the second quantization formalism is 

\begin{align}
\rho^{(1)} = \text{Tr}^{(1)} \ket{\Phi}\bra{\Phi} = \frac{\sum_k a_{\psi_k} \ket{\Phi} \bra{\Phi} a_{\psi_k}^\dagger}{\braket{\Phi|\hat{\mathbf{n}}|\Phi }}
\end{align}  
Here $ \hat{\mathbf{n}} = \sum_k  a_{\psi_k}^\dagger a_{\psi_k} $ is the total number operator. The details are given in appendix \ref{part_sec_quant}.
\section{Anyons}
It is well-known that, in relativistic quantum field theory, the spin-statistics theorem\cite{pauli1940connection} 
dictates that bosonic fields satisfy canonical commutation relations, while fermionic fields satisfy anti-commutation relations.  
In nonrelativistic quantum mechanics, one 
mimics the quantum field theoretic ideas through second quantization which directly
yields multi-particle wave functions of indistinguishable particles with appropriate 
symmetry properties. In particular, particles with (half-)integer spin have wave functions which are (anti-)symmetric under the exchange of any two particles. 

In contrast, the Symmetrization Postulate \cite{messiah} accomplishes  this objective by attaching labels to the particles, as if they were distinguishable, and  (anti-)symmetrizing the product wave function with respect to these labels.  But, labelling indistinguishable particles is intrinsically contradictory. So, it is desirable to look beyond this ad hoc prescription.

In a seminal paper, Leinaas and Myrheim \cite{leinaas1977theory} trace the origin of the Symmetrization Postulate to the non-trivial topology of the underlying classical configuration space of indistinguishable particles. As a spin-off of this insight,
they show that, in low dimensions, it is possible to have objects which are more general than bosons and fermions. These are called anyons.
In what follows, we briefly summarise the 
Leinaas-Myrheim method that leads to anyons.  

Let us consider a system of $ N $ spin-less particles in $d$ dimensions. Let  $X = {\mathbf R}^d$ be  the configuration space of a single particle. If the particles are distinguishable, the configuration space of the system is $\mathcal{X}_N = X^N$ where $ X^N $ denotes an $ N $ - fold tensor product of the single-particle space $ X $. A point in the space $\mathbf{x} = \lr{x_1,x_2,...,x_N}$ represents a physical configuration of the system.

If the $ N $ particles are indistinguishable,  the configuration space is $\mathcal{Y}_N = (X^N-D)/S_N$ where $S_N$ is the permutation group on $N$ elements.  It ensures that the points $\bold{x} = \lr{x_1,x_2,...,x_N}$ and $\bold{x'} = \lr{x_{P(1)},x_{P(2)},...,x_{P(N)}}$ which represent the same physical configuration are identified. Here $P$ represents an arbitrary permutation.  $D$ represents the set of 
singular points which are unaffected by the 
identifications. 

In the above, the description is entirely classical. The idea is that since the identifications have been made already at the level of the classical configuration space, the restrictions on quantum states would follow without the ad hoc need to invoke the symmetrization postulate. For 
particles with spin, one continues to define the 
configuration space as above, with the minor 
modification that at each point in ${\cal Y}_N$  we erect a spinor space. The spin observables
act as operators on this spinor space. We refer the reader to \cite{leinaas1977theory} for further details.  

In the above formalism, the quantum mechanical wave function of the system is determined by the one-dimensional unitary representations of the fundamental group $\pi_1(\mathcal{Y}_N)$ of the  configuration space.  For the case of  indistinguishable particles, this turns out to be the permutation group  in dimensions $d\geq 3$, whose lowest dimensional irreducible representations allow only bosons and fermions.  In two dimensions, the fundamental group of the system is $\pi_1\rl{\mathcal{Y}_{N}} = B_N$, where $B_N$ is the braid group on $N$ strings, whose one dimensional unitary representations allow the wave function to pick up a  phase $e^{i \theta}$, where $\theta$ is a real parameter, under an exchange. This is the underlying reason for the possibility of having anyons in 
low dimensions. \\

\subsection*{Indistinguishable Particles On the Real Line}

In the case of indistinguishable particles on a real line, it is not possible to perform an exchange without taking the
particles  through each other:  an exchange gets inextricably linked with scattering. It is neverthless possible to define quantum statistics by following the 
Leinaas-Myrheim prescription, as shown below in the 
specific example of two indistinguishable particles on a 
real line. If $x_1$ and $x_2$ are the positions of the particles,  we observe that the points $\bold{x} = \rl{x_1,x_2}$ and  $\bold{x'} = \rl{x_2,x_1}$ represent the same configuration, and hence need to be identified. The identification is done by folding the $(x_1x_2)$ plane along
the line $x_1=x_2$ which represents the singular points. Without loss of generality, we choose to work with the half plane $x_1<x_2$.  The problem can be solved by prescribing appropriate boundary conditions along
the diagonal. \\

We choose the free particle Hamiltonian for the system, also studied by Posske et al \cite{posske2017second}, 
\begin{align}
H = -\frac{1}{2}\rl{\frac{\partial^2}{\partial x_1^2}+\frac{\partial^2}{\partial x_2^2}}
\end{align}
where we use the units $\hbar=c=1$ and set mass equal
to one. To ensure that particles remain bounded in the region $x_1<x_2$, we impose the boundary condition that the normal component of the probability current vanishes at the boundary. That is,
\begin{equation}
\rl{\psi^*(\bold{x}) \rl{-\frac{\partial}{\partial x_1} + \frac{\partial}{\partial x_2}} \psi(\bold{x}) - \psi(\bold{x}) \rl{-\frac{\partial}{\partial x_1} + \frac{\partial}{\partial x_2}} \psi^*(\bold{x})}\bigg \rvert_{x_1=x_2} =0
\end{equation}
Note that above equation also ensures self-adjointness of the Hamiltonian. The general solution of the above equation is given by,
\begin{equation}
\rl{-\frac{\partial}{\partial x_1} + \frac{\partial}{\partial x_2}} \psi(\bold{x}) \bigg \rvert_{x_1=x_2}= \eta  \psi(\bold{x}) \bigg \rvert_{x_1=x_2} 
\end{equation}
where $\eta $ is a real parameter. The eigenstates of the Hamiltonian are
\begin{equation}
\psi(\bold{x}) = e^{i \left(k_1 x_1+k_2 x_2\right)} +e^{-i \rl{\phi_\eta \rl{k_{2}-k_{1}}}} e^{i
   \left(k_2 x_1+k_1 x_2\right)}
\end{equation}
where, \begin{align*}
	\phi_\eta \rl{k_{2}-k_{1}} = 2 \tan^{-1} \rl{\frac{\eta}{k_{2}-k_{1}}}
\end{align*}
Note that $\eta = 0$ and $\eta =\infty$  correspond to 
Neumann and Dirichlet boundary conditions respectively
on the diagonal {\it i.e.} the set of coincident points $x_1=x_2$. The former gives a symmetric wave function, while the latter gives an antisymmetric wave function 
which also enforces the Pauli Exclusion Principle.  
Arbitrary values of $\eta$ correspond to Robin
boundary conditions, with the corresponding wave functions being neither symmetric nor antisymmetric. These are, by definition, one-dimensional anyons.  

For $\eta <0$, it is easy to see that the system admits one
bound state.\footnote{The Hamitonian, despite its appearance,  it is not positive definite because of the boundary. This is what allows for the existence of a bound state.} This follows from the requirement that the 
wave function is well-behaved at $\pm\infty$, which in turn implies that the momentum of the centre of mass coordinate is purely real, and the momentum of the 
relative coordinate is purely imaginary.

We mention in passing that for the case of three or more particles, there are several diagonals corresponding to 
coincident points; but the Robin boundary conditions can be generalized in a straightforward manner as shown in the next subsection.

\subsubsection*{$N$ particles on the real line}
In the case of $N$ identical particles on a real line the configuration space can be constructed in a similar way and is chosen to be the region where $\mathcal{R} = \lbrace\mathbf{x} | x_1<x_2<x_3<...<x_N \rbrace$. The Hamiltonian is again the free particle Hamiltonian
\begin{align}
H =-\frac{1}{2} \sum_{j=1}^N \frac{\partial^2}{\partial x_j^2} 
\end{align}
and the Robin boundary conditions  are
\begin{align}
\rl{\frac{\partial}{\partial x_{j+1}} - \frac{\partial}{\partial x_j}} \psi(\bold{x}) \bigg \rvert_{x_{j+1}=x_j}= \eta  \psi(\bold{x}) \bigg \rvert_{x_{j+1}=x_j} 
\end{align}
The corresponding anyonic wave functions are obtained by 
solving the Schrodinger equation for which we employ the ansatz $\psi\lr{\bold{x}}= \int_{\bold{k}  \in \mathbb{C}^n} d\bold{k} \ \alpha\rl{\bold{k}} e^{i \bold{k} \bold{x}}$. The coefficients $\alpha\rl{k}$ satisfy,
\begin{equation}
\alpha\rl{\bold{k}} =\begin{cases} e^{-i \rl{\phi_\eta \rl{k_{j+1}-k_{j}}}}\alpha\rl{P_j \bold{k}} \ \ \text{if} \ \ k_{j+1}-k_j \neq i 
\eta \\ 0 \ \ \text{if} \ \  k_{j+1}-k_j = i \eta \end{cases}
\end{equation}
where an elementary permutation $P_j$  permutes the 
$ j $th and $ (j+1)$th elements and 
\begin{align}
\phi_\eta \rl{k_{j+1}-k_{j}} = 2 \tan^{-1} \rl{\frac{\eta}{k_{j+1}-k_{j}}}
\end{align}
The connection between the coefficients can be written as follows
\begin{align}
\alpha\rl{\bold{k}} = e^{i \phi_\eta^P\rl{\bold{k}}} \alpha\rl{P \bold{k}}
\end{align}
where $P = P_{j_1}.....P_{j_r}$ represents the minimum 
number of elementary permutations required to reach 
a given permutation.  
\begin{align*}
\phi_\eta^P \rl{\bold{k} }= \sum_{i=1}^r \phi_\eta \left[\rl{P_{j_1}.....P_{j_i} \bold k}_{j_i} -\rl{P_{j_1}.....P_{j_i} \bold k}_{j_{i+1}} \right].
\end{align*}
 The basis functions are of the form $\psi_\bold{k} \rl{\bold{x}} \varpropto  \sum_{P \in S_n} e^{i \phi_\eta^P\rl{\bold{k}}} e^{i \rl{P \bold k} \bold{x}}$. 
As in the two-particle case, only special values of 
$\bold{k} $ are permitted when $\eta <0$.  In contrast to the two-particle case, however, we can have bound states with different number of particles.  
\subsection*{Second quantization}
As already mentioned in the Introduction, we find it useful
to recast the above results in the language of second quantization, as was done in \cite{second}.  We use the following generalised $\eta$-dependent algebra for the second quantized creation operator $\Psi^\dagger(x)$,  and annihilation operator $\Psi(x)$ of the anyon fields
\begin{align}\label{realspace}
\left[ \Psi(x),\Psi^\dagger(y) \right] &= \delta(x-y)- 2 \eta  \int_{0}^\infty dz \  e^{-z \eta} \Psi^{\dagger}(y-z) \Psi(x-z) \no \\
\left[ \Psi^\dagger(x),\Psi^\dagger(y) \right] &=-2 \eta  \int_{0}^\infty dz \  e^{-z \eta} \Psi^{\dagger}(y+z) \Psi^\dagger(x-z).
\end{align}
Note that this algebra reduces to the standard bosonic and fermionic limits for $\eta \to 0$ and $\eta \to \infty$ respectively. Also note that this algebra is slightly different from the one presented in \cite{second}. As shown in appendix \ref{derreal}, the above equations can be derived starting from the corresponding algebra for the creation and annihilation operators for momentum states, related to 
the second quantized fields through the usual relations  $\Psi^\dagger(x) = \frac{1}{\sqrt{2\pi}}\int_{-\infty}^\infty dk e^{i k x} a_k^\dagger $.  

The following commutators involving the number operator $\hat N$ defined in the usual manner as $\hat{N} = \int_{-\infty}^{\infty} dx \  \Psi^\dagger(x) \Psi(x)$, can be derived in a 
straightforward manner as shown in appendix \ref{numbop}
\begin{align}
\begin{aligned}
\left[ \hat{N}, \Psi^\dagger(y)   \right] & = \Psi^\dagger(y)  \\
\left[ \hat{N}, \Psi(y)   \right]  &=  - \Psi(y)  
\end{aligned}
\end{align}
Thus, although the algebra for the anyonic fields is 
more complicated than the bosonic and fermionic 
cases, the number operator can be defined in the 
usual fashion, and satisfies the standard commutation
relation with the second quantized fields. This allows 
us to interpret the matrix elements of the fields 
in the number operator basis as operators which 
transform multiparticle wave
functions into other wave functions with  more or fewer number of particles as explained by Fock
\cite{faddeev2004va}. In appendix \ref{checkalgebra}, we explicitly verify that the 
modified algebra satisfies the conditions derived by Fock.

\section{Entropy of Two Identical Particles}
We consider two indistinguishable particles on the real line. We assume that the statistics parameter $\eta$ is non-negative, so that the particles are anyons. Note that the bosonic and fermionic limits can be retrieved from the general case as special cases. 

The field operator $\Psi^\dagger (x)$ acting on the vacuum  creates a particle localised at $x$. Rather than dealing with these localised states, it is convenient for 
our purposes  to work with smeared fields defined as follows: $\Psi_f^\dagger = \int_{-\infty}^\infty dx  \ f(x) \Psi^\dagger\rl{x}  $, where $f(x) \in \mathcal{S} (\mathbb{R})$, is a function in the Schwartz space \cite{reed2012methods}. The algebra of the smeared fields is readily obtained to be 
\begin{align}\label{alg}
\left[ \Psi_{f} ,\Psi_{g} ^\dagger \right] &= \braket{f|g} - 2 \eta \int_{0}^\infty dz  \int_{-\infty}^\infty \int_{-\infty}^\infty dxdy f^*(x) g(y)  e^{-z \eta} \Psi^{\dagger}(y-z) \Psi(x-z) \no \\
\left[ \Psi^\dagger_{f},\Psi_{g} ^\dagger \right] &=- 2 \eta \int_{0}^\infty dz  \int_{-\infty}^\infty \int_{-\infty}^\infty dxdy f^*(x) g(y)  e^{-z \eta} \Psi^{\dagger}(y+z) \Psi^\dagger(x-z)
\end{align}
where the inner product $\braket{f|g} = \int_{-\infty}^{\infty} dx \  f^*(x)g(x)$. We use the following notation to denote the states $\ket{f} \equiv \Psi_f^\dagger \ket{0}$. If we choose a set of orthonormal functions $\lbrace f_n\rl{x}  \rbrace$, the corresponding set of states $\lbrace \ket{f_n} \rbrace$ will form a basis for the single-particle Hilbert space. For our purpose we chose $f_n(x) = h_n(x)$, where $h_n(x) = \frac{1}{\sqrt{\sqrt{\pi} 2^n n! }} H_n\rl{x} e^{-\frac{x^2}{2}} $ is $n$-th eigenstate of the harmonic oscillator.

Let the two-particle state be
\begin{align}\label{state}
\ket{\Phi_{j,i}} \equiv \frac{1}{\mathcal{N}}\Psi_{h_j}^\dagger \Psi_{h_i}^\dagger \ket{0} 
\end{align}
Here $ \mathcal{N}  =\bra{0} \Psi_{h_i} \Psi_{h_j}  \Psi_{h_j}^\dagger \Psi_{h_i}^\dagger \ket{0} $ is the normalization constant.
We use the one-particle basis $\lbrace \ket{h_n} \rbrace$   as the basis to calculate both the partial trace and the eigenvalues of the reduced density matrix. The one-particle reduced density matrix $\rho^{(1)}$ is obtained from the two-particle state as follows
\begin{align*}
\rho^{(1)} = \frac{\sum_{k=0}^{\infty} \Psi_{h_k} \ket{\Phi_{j,i}}\bra{\Phi_{j,i}}  \Psi_{h_k}^\dagger }{\bra{\Phi_{j,i}}\hat{\mathbf{n}}  \ket{\Phi_{j,i}}}
\end{align*} 
where $ \hat{\mathbf{n}} =  \sum_{k=0}^{\infty}  \Psi_{h_k}^\dagger \Psi_{h_k} $ is the total number operator. A matrix element of the reduced density matrix is given by
\begin{align*}
\rho^{(1)}_{m,n} = \braket{h_m| \rho^{(1)}|h_n}  =  \frac{\sum_{k=0}^{\infty} \bra{0}\Psi_{h_m}  \Psi_{h_k} \ket{\Phi_{j,i}}\bra{\Phi_{j,i}} \Psi_{h_k}^\dagger \Psi_{h_n} \ket{0} }{\bra{\Phi_{j,i}}\hat{\mathbf n} \ket{\Phi_{j,i}}}
\end{align*}
The expressions for the matrix element can be obtained analytically. They are given by an infinite series involving parabolic cylinder functions. They depend on $\eta$. The detailed calculations are given in appendix \ref{redmat}. 

Since the expressions for the reduced density matrix are
cumbersome, we resort to calculating the eigenvalues numerically, by using the formula
\begin{align*}
\sum_{m=0}^{\infty} \rho^{(1)}_{m,n}  g\rl{n} = \lambda_n g\rl{m}
\end{align*}
where $\lambda_n$ is an eigenvalue. The von Neumann entropy is then given by the usual formula
\begin{align*}
S\rl{\rho^{1}} =- \text{Tr} \rl{\rho^{1} \log \rl{\rho^{1}}} = -\sum_i \lambda_i \log \rl{\lambda_i}
\end{align*}

The dependence of the von Neumann entropy on the statistics parameter $\eta$ is plotted in the following 
figures for different choices of the initial two-particle state.
\begin{figure}[H]\label{plot1}
\centering
	\includegraphics[scale=0.7]{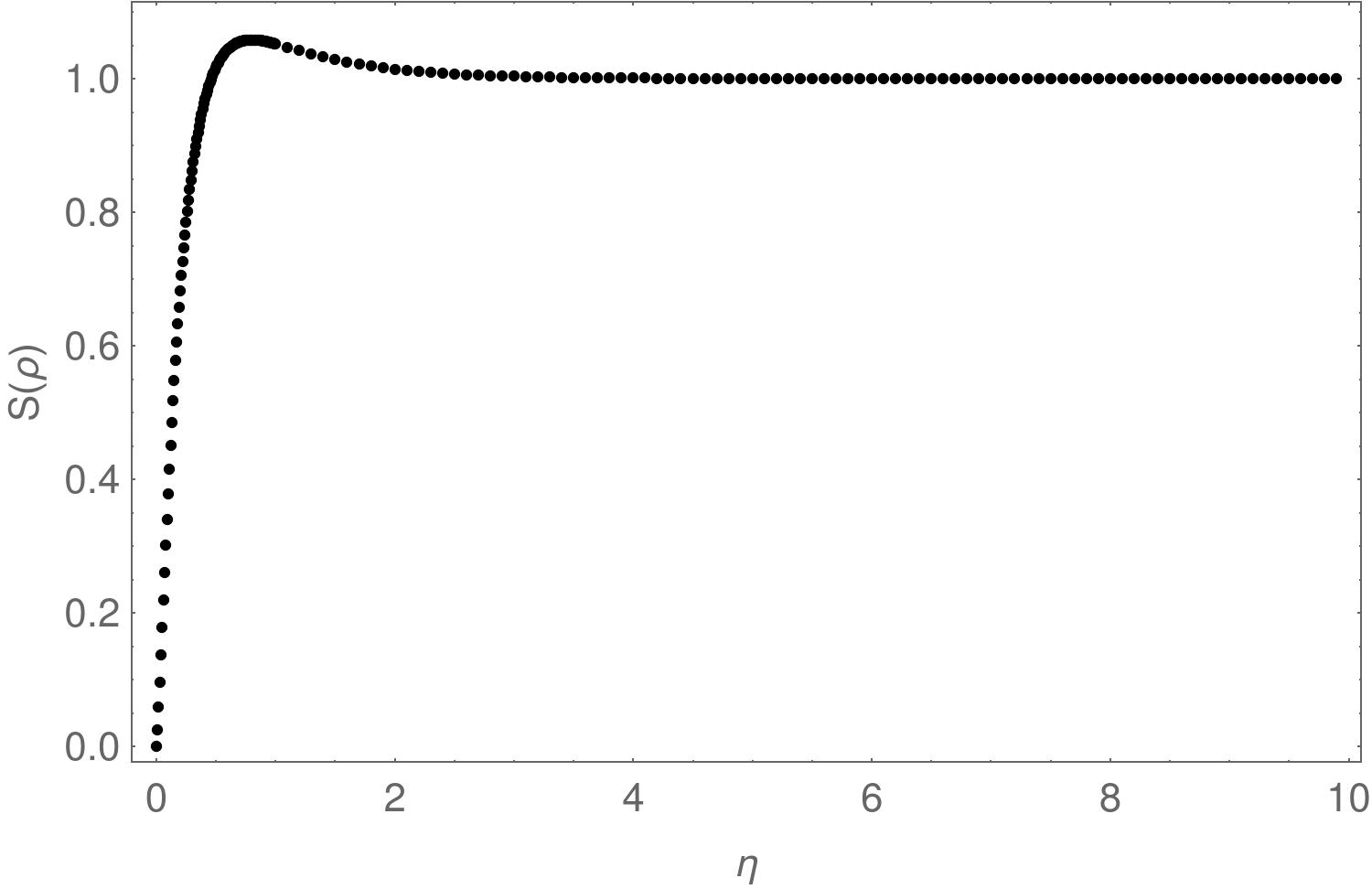}
\caption{Plot of entropy vs statistics parameter $\eta$ for the initial two-particle state $\ket{\Phi_{0,0}}$.}
\label{fig:plot1}
\end{figure}
In the above plot, the two-particle state is taken to be $\ket{\Phi_{0,0}}$. It is 
worth noting that for $\eta =0$, both the particles are in 
the same state. The entropy is zero, consistent with what
is expected of bosons. Note, however, that this plot is not 
valid in the fermionic limit $\eta\to\infty$, because the state $\ket{\Phi_{0,0}} $ identically vanishes as can be easily seen from equations\ref{alg}, \ref{state}.

\begin{figure}[H]\label{plot2}
	\centering
	\includegraphics[scale=0.7]{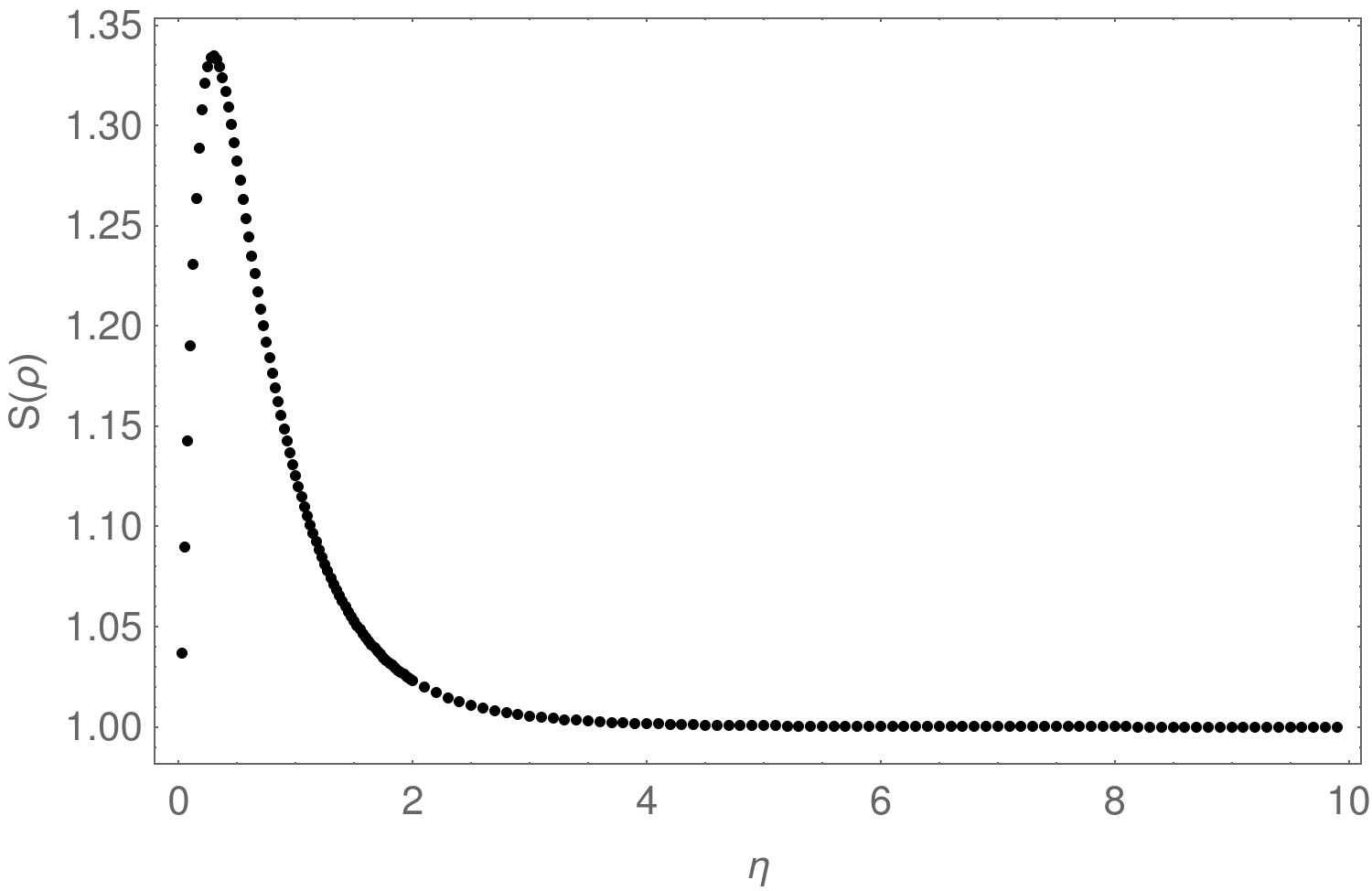}
	\caption{Plot of entropy vs statistics parameter $\eta$ for the initial two-particle state $\ket{\Phi_{1,0}}$.}
	\label{fig:plot2}
\end{figure}
In the above plot, the two-particle state is taken to be $\ket{\Phi_{1,0}}$. In this 
case, it worth noting that for both $\eta =0$ and $\eta \to \infty$, the entropy is equal to unity. 

In order to get a better insight into what the above plots mean, it is useful to compare our results with  \cite{franco2016quantum}. Lo Franco and Compagno consider a model of two indistinguishable qubits in an asymmetric double-well potential. In particular, they study the spin correlations between the qubits in the 
same spatially localised state, namely the left trough. It is important to note that the potential acts as a crutch to produce various states for the qubits, namely, states which are localised either on the left side, or the right side, or those which are in a superposition of the left and right 
sides. Once a state is specified, only the finite-dimensional Hilbert spaces associated with the qubits play a role. For example, they show that when both the qubits are localised in the left well, the state $\ket{L \uparrow, L\uparrow}$ is not entangled, whereas, the state $\ket{L\uparrow, L \downarrow}$ is maximally entangled analogous to the Bell state for distinguishable qubits. In arriving at this result the one-particle basis used is finite-dimensional,
because only the spin degrees of freedom of the qubits are considered.

In our model, the states $\ket{\Phi_{0,0}}$ 
and $\ket{\Phi_{1,0}}$ are analogous to the states $\ket{L\uparrow, L\uparrow}$ and  $\ket{L\uparrow, L\downarrow}$. But there are 
crucial differences. The states in our model represent not two indistinguishable qubits, but two indistinguishable particles. This has important ramifications. 

First, the entropy need not be bounded by unity.  Second, it depends on the statistics parameter $\eta$. That is what is displayed in the above plots.  From these one can read off the approximate values of the entropy obtained using numerical analysis for any given value of $\eta$ .  

It is interesting to note that in spite of these differences our results agree with \cite{franco2016quantum} in the limiting 
cases of $\eta\to 0$ and $\eta\to\infty$, corresponding to bosons and fermions respectively. To understand this one has to look at the non-vanishing eigenvalues of the 
reduced density matrices. However, one has to remember that the two systems are really physically very different. A subtle point to note
is that, as already pointed out, the states $\ket{\Phi_{0,0}}$ and $\ket{\Phi_{1,0}}$ are analogous to the states $\ket{L\uparrow, L\uparrow}$ and  $\ket{L\uparrow, L\downarrow}$ respectively. To be more precise, as $\eta\to 0$, namely the bosonic limit, the state $\ket{\Phi_{1,0}}$ is entangled, so is the bosonic state $\ket{L\uparrow, L\downarrow}$. As $\eta\to\infty$, namely the fermionic limit, the state $\ket{\Phi_{1,0}}$ is entangled, so is the fermionic state $\ket{L\uparrow, L\downarrow}$. As $\eta\to 0$, the state $\ket{\Phi_{0,0}}$ is entangled, so is the bosonic state  $\ket{L\uparrow, L\uparrow}$. Finally as $\eta\to\infty$,  the state $\ket{\Phi_{0,0}}$ vanishes 
as already explained, and the fermionic state $\ket{L\uparrow, L\uparrow}$ is identically zero due to Pauli's exclusion principle. Hence qualitatively, the two 
systems appear to be identical in these limits if we formally identify the spin degrees of freedom 
of the qubit with the two levels labeling the 
$\Phi_{j,i}$.
 
The other results that Lo Franco and Compagno obtain regarding non-local entanglement use superpositions of states localised in the left and right wells, and are beyond the scope of the present work.  

\section{Conclusions}
The problem of studying the entanglement between indistinguishable particles in quantum mechanics is 
tricky.  A naive usage of the usual measures like the Schmidt rank and the von Neumann entropy leads to 
wrong results. 

A way to bypass these problems, restricted to bosons 
and fermions, was developed by Lo Franco and Compagno \cite{franco2016quantum} by using ideas coming from 
information theory. 

In this paper we use their results, in the second quantized 
formulation,  to study the entanglement between two one-dimensional anyons.  The generalised algebra of one-dimensional anyons obtained from a second quantization of the Leinaas-Myrheim model \cite{second} plays a crucial role in our analysis. 

We succeed in obtaining both qualitative and approximately quantitative results for the dependence of the von Neumann entropy on the statistics parameter.   

The calculations presented in this paper are readily 
generalizable to studying entanglement between two clusters of anyons with an arbitrary number of particles. Other one-dimensional models admitting
anyonic statistics like indistinguishable particles on a ring
and the Calogero model are also worth investigating.

The most interesting problem will, of course, be to investigate the entanglement between anyons in two dimensions, both in the abelian and non-abelian cases,
because of their direct relevance to topological quantum computation. 

We hope to address these questions in our future work. 
\section{Acknowledgments}
We thank Alok Laddha  for discussions and the referees 
for many useful suggestions.  This work is partially supported by a grant to CMI from the Infosys Foundation.

\appendix
\section{The reduced density matrix in the second quantization formalism}\label{part_sec_quant}
In the second quantization language the $ N $-particle state $ \ket{\Phi} $ is obtained by acting with a suitable combination of creation operators on the vacuum state. Let the set of states $ \lbrace \ket{\psi_k} \equiv a^\dagger_{\psi_k} \ket{0}  \rbrace \ $ form a basis for the single particle Hilbert space. In analogy with equation \ref{projsingle}, the state $ \ket{\phi_k} $ is defined in the second quantization formalism as follows
\begin{align*}
\ket	{\phi_k} = \frac{a_{\psi_k} \ket{\Phi}}{\sqrt{\braket{\Phi|a_{\psi_k}^\dagger a_{\psi_k}| \Phi }}}
\end{align*}
The corresponding probabilities are 
\begin{align*}
p_k = \frac{\braket{\Phi|a_{\psi_k}^\dagger a_{\psi_k}| \Phi }}{\braket{\Phi|\hat{\mathbf{n}}| \Phi }}
\end{align*}
where $ \hat{\mathbf{n}} = \sum_k  a_{\psi_k}^\dagger a_{\psi_k} $ is the total number operator. Then, the one-particle reduced density matrix is
 \begin{align*}
 \rho^{(1)} = \frac{\sum_k a_{\psi_k} \ket{\Phi} \bra{\Phi} a_{\psi_k}^\dagger}{\braket{\Phi|\hat{\mathbf{n}}|\Phi }}
 \end{align*}  
 \section{Real space algebra}
\subsection{Derivation of the real space algebra}\label{derreal}
The algebra of creation and annihilation operators of momentum states obtained in \cite{second} is
\begin{align}
a_p^\dagger  a_q^\dagger &= e^{i \phi_\eta (p-q)} a_q^\dagger a_p^\dagger  \no \\
a_p a_q^\dagger &= e^{-i \phi_\eta (p-q)} a_q^\dagger a_p + \delta(p-q)
\end{align}
where the phase $e^{i \phi_\eta (p-q)} = \frac{p-q + i \eta}{p-q - i \eta}$. The above relations may be rewritten 
in a slightly modified way as follows
\begin{align}
a_p^\dagger  a_q^\dagger &= \rl{\frac{p-q + i \eta}{p-q - i \eta}} a_q^\dagger a_p^\dagger  \no \\
a_p a_q^\dagger &= \rl{\frac{p-q - i \eta}{p-q + i \eta}} a_q^\dagger a_p + \delta(p-q)
\end{align}
Note that the creation and annihilation operators for the  momentum states are related to 
the second quantized fields through the relations 
\[\Psi^\dagger(x) = \frac{1}{\sqrt{2\pi}}\int_{-\infty}^\infty dk e^{i k x} a_k^\dagger .
\]
 
To obtain the algebra of field operators we calculate the commutator between field operators
\begin{align*}
\Psi(x) \Psi^\dagger(y) - \Psi^\dagger(y) \Psi(x) &=  \int_{-\infty}^\infty  dp dq e^{-ipx+iqy} \rl{a_p a_q^\dagger - a_q^\dagger a_p }
\end{align*}
Substituting for $a_p a_q^\dagger$ from the algebra of creation and annihilation operators for momentum states,
\begin{align*}
\Psi(x) \Psi^\dagger(y) & - \Psi^\dagger(y) \Psi(x) \\ &=  \frac{1}{2\pi}\int_{-\infty}^\infty  dp dq e^{-ipx+iqy} a_q^\dagger a_p \rl{  \rl{\frac{p-q - i \eta}{p-q + i \eta}}-1 } + \delta(p-q)\\
&= \delta(x-y) - \frac{ \eta}{\pi} \int_{-\infty}^\infty  dp dq e^{-ipx+iqy} a_q^\dagger a_p \rl{  \frac{1}{-ip+iq +  \eta}} \\
&= \delta(x-y) -\frac{ \eta}{\pi}   \int_0^{\infty} dz e^{-z \eta}  \int_{-\infty}^\infty  dp dq e^{-ip(x-z)+iq(y-z)} a_q^\dagger a_p\\
&= \delta(x-y) -2  \eta  \int_0^{\infty} dz e^{-z \eta} \Psi^{\dagger}(y-z) \Psi(x-z)
\end{align*}
Similarly, if we look at the commutator $ \left[ \Psi^\dagger(x) ,\Psi^\dagger(y) \right] $, we obtain 
\begin{align*}
\Psi^\dagger(x) \Psi^\dagger(y) - \Psi^\dagger(y) \Psi^\dagger(x) &=  \int_{-\infty}^\infty  dp dq e^{ipx+iqy} \rl{a_p^\dagger a_q^\dagger - a_q^\dagger a_p^\dagger }
\end{align*}
Substituting for  $a_p^\dagger a_q^\dagger$ ,
\begin{align*}
\Psi^\dagger(x) \Psi^\dagger(y) &- \Psi^\dagger(y) \Psi^\dagger(x) \\ &=  \frac{1}{2\pi} \int_{-\infty}^\infty  dp dq e^{ipx+iqy} a_q^\dagger a_p^\dagger  \rl{ \rl{\frac{p-q + i \eta}{p-q - i \eta}}  - 1 } \\
& = -\frac{ \eta}{\pi}  \int_{-\infty}^\infty  dp dq e^{ipx+iqy} a_q^\dagger a_p^\dagger \int_0^\infty dz e^{-z(ip-iq + \eta)} \\ 
& = -\frac{ \eta}{\pi}  \int_0^\infty dz e^{-z \eta} \int_{-\infty}^\infty  dp dq e^{ip(x-z)+iq(y+z)} a_q^\dagger a_p^\dagger  \\
& = -2 \eta \int_0^\infty dz e^{-z \eta} \Psi^\dagger(y+z) \Psi^\dagger(x-z)
\end{align*}
Instead if we substitute for $a_q^\dagger a_p^\dagger$,
\begin{align*}
\Psi^\dagger(x) \Psi^\dagger(y) - \Psi^\dagger(y) \Psi^\dagger(x) &=  \int_{-\infty}^\infty  dp dq e^{ipx+iqy} \rl{a_p^\dagger a_q^\dagger - a_q^\dagger a_p^\dagger }\\
&=  \int_{-\infty}^\infty  dp dq e^{ipx+iqy} a_p^\dagger a_q^\dagger \rl{\frac{2 i \eta}{p-q + i \eta}}\\
&=2 \eta  \int_{-\infty}^\infty  dp dq e^{ipx+iqy} a_p^\dagger  a_q^\dagger  \int_0^\infty dz e^{-z (-ip+iq +  \eta)} \\
& = 2 \eta \int_0^\infty dz e^{-z \eta} \Psi^\dagger(x+z) \Psi^\dagger(y-z)
\end{align*}

\subsection{Commutation relations involving number operator}\label{numbop}
The number operators is $\hat{N} = \int_{-\infty}^{\infty} dx \  \Psi^\dagger(x) \Psi(x)$. We calculate the commutator between the number operator and the field theoretic anyon creation operator by substituting in terms of momentum space operators as follows 
\begin{align*}
\left[ \hat{N}, \Psi^\dagger(y)   \right] &=   \int_{-\infty}^{\infty} dx  \left[ \Psi^\dagger(x) \Psi(x), \Psi^\dagger(y)   \right] \\
&= \frac{1}{\rl{2 \pi}^{\frac{3}{2}}}  \int_{-\infty}^{\infty} dx \int_{-\infty}^{\infty} dpdqdr \ e^{ipx-iqx+iry} \rl{a_p^\dagger a_q a_r^\dagger - a_r^\dagger a_p^\dagger a_q }  \\
&=   \frac{1}{\rl{2 \pi}^{\frac{3}{2}}} \int_{-\infty}^{\infty} dpdr \ e^{iry} \rl{a_p^\dagger a_p a_r^\dagger - a_r^\dagger a_p^\dagger a_p }  \\
&=   \frac{1}{\rl{2 \pi}^{\frac{3}{2}}} \int_{-\infty}^{\infty} dpdr \ e^{iry} \rl{a_p^\dagger \rl{e^{-i \phi_\eta (p-r)} a_r^\dagger a_p + \delta(p-r)} - a_r^\dagger a_p^\dagger a_p }  \\
&=   \frac{1}{\rl{2 \pi}^{\frac{3}{2}}} \int_{-\infty}^{\infty} dpdr \ e^{iry} \rl{\delta(p-r) +  e^{-i \phi_\eta (p-r)}e^{i \phi_\eta (p-r)} a_r^\dagger a_p^\dagger a_p  - a_r^\dagger a_p^\dagger a_p }  \\
& =  \Psi^\dagger(y) 
\end{align*}
The corresponding result for the annihilation operator is
\begin{align*}
\left[ \hat{N}, \Psi(y)   \right]  &=  - \Psi(y)  
\end{align*}
The same results can be obtained using the real space operator algebra as shown below
\begin{align*}
\left[ \hat{N}, \Psi^\dagger(y)   \right] 
& = \int_{-\infty}^{\infty} dx \rl{  \Psi^\dagger(x) \Psi(x) \Psi^\dagger(y) - \Psi^\dagger(y) \Psi^\dagger(x) \Psi(x)  }  \\
&=   \int_{-\infty}^{\infty} dx \bigg(  \Psi^\dagger(x) \bigg( \Psi^\dagger(y) \Psi(x) + \delta(x-y) \\&-2  \eta  \int_0^{\infty} dz e^{-z \eta} \Psi^{\dagger}(y-z) \Psi(x-z) \bigg) - \Psi^\dagger(y) \Psi^\dagger(x) \Psi(x)  \bigg)  \\
&=  \Psi^\dagger(y) \\&+ \int_{-\infty}^{\infty} dx \bigg(\bigg(   \rl{ \Psi^\dagger(y)\Psi^\dagger(x) + 2 \eta \int_0^\infty dz e^{-z \eta} \Psi^\dagger(x+z) \Psi^\dagger(y-z)} \Psi(x) \\ & \quad -2  \eta  \int_0^{\infty} dz e^{-z \eta} \Psi^\dagger(x) \Psi^{\dagger}(y-z) \Psi(x-z) \bigg)  - \Psi^\dagger(y) \Psi^\dagger(x) \Psi(x) \bigg)    \\
&=  \Psi^\dagger(y) 
\end{align*}
With a similar calculation, we can obtain the commutator of the number operator with field theoretic anyon annihilation operator.
\subsection{Check on the algebra}\label{checkalgebra}

The symmetrization postulate for multiparticle wave functions in the first quantized formalism has an 
intimate connection with the algebra of creation and 
annihilation operators in the second quantized formalism.
This was clearly explained in very general terms by Fock
for the case of bosons and fermions in \cite{faddeev2004va}. In this appendix we verify the 
consistency of the anyonic algebra we use along similar lines. 
 
The field operator $ \Psi(x) $ acts on the sequence of functions 
\begin{align}
\begin{pmatrix}
\text{const.}\\\psi(x_1)\\ \psi(x_1,x_2) \\ \psi(x_1,x_2,x_3) \\ .......\\
\end{pmatrix}
\end{align}
as follows
\begin{align}
\Psi(x)\begin{pmatrix}
\text{const.}\\\psi(x_1)\\ \psi(x_1,x_2) \\ \psi(x_1,x_2,x_3) \\ .......\\
\end{pmatrix}
=
\begin{pmatrix}
\psi(x)\\\sqrt{2} \psi(x,x_1)\\ \sqrt{3} \psi(x,x_1,x_2) \\  \sqrt{4} \psi(x,x_1,x_2,x_3) \\ .......\\
\end{pmatrix}
\end{align}
where the functions $ \psi(x_1), \psi(x_1,x_2) , \psi(x_1,x_2,x_3).... $ are interpreted as Schrodinger wave functions\cite{faddeev2004va}. \\

Applying the operator $ \Psi(x') \Psi(x) $ on the sequence of functions, we obtain
\begin{align}\label{eq1}
\Psi(x')\Psi(x)\begin{pmatrix}
\text{const.}\\\psi(x_1)\\ \psi(x_1,x_2) \\ \psi(x_1,x_2,x_3) \\ .......\\
\end{pmatrix}
=
\begin{pmatrix}
\sqrt{2.1}\psi(x,x')\\\sqrt{3.2} \psi(x,x',x_1)\\ \sqrt{4.3} \psi(x,x',x_1,x_2) \\  \sqrt{5.4} \psi(x,x',x_1,x_2,x_3) \\ .......\\
\end{pmatrix}.
\end{align}

Similarly, applying the operator $ \Psi(x) \Psi(x') $ on the sequence of functions, we get
\begin{align}\label{eq2}
\Psi(x)\Psi(x')\begin{pmatrix}
\text{const.}\\\psi(x_1)\\ \psi(x_1,x_2) \\ \psi(x_1,x_2,x_3) \\ .......\\
\end{pmatrix}
=
\begin{pmatrix}
\sqrt{2.1}\psi(x',x)\\\sqrt{3.2} \psi(x',x,x_1)\\ \sqrt{4.3} \psi(x',x,x_1,x_2) \\  \sqrt{5.4} \psi(x',x,x_1,x_2,x_3) \\ .......\\
\end{pmatrix}
\end{align}
In the case of bosons, the right hand side of Eq.\ref{eq1} and Eq.\ref{eq2} are the same because the bosonic wave function is symmetric under the exchange of any pair of coordinates. This implies that the field operators $ \Psi(x) $ and $ \Psi(x') $ commute with each other.  In the case of fermions, using the same argument and by noting that the fermionic wave functions are anti-symmetric, one can obtain the usual anti-commutation relation between $ \Psi(x) $ and $ \Psi(x') $. 

In our case the field operators satisfy the following algebra
\begin{align*}
\left[ \Psi(x),\Psi^\dagger(y) \right] &= \delta(x-y)- 2 \eta  \int_{0}^\infty dz \  e^{-z \eta} \Psi^{\dagger}(y-z) \Psi(x-z) \no \\
\left[ \Psi^\dagger(x),\Psi^\dagger(y) \right] &=-2 \eta  \int_{0}^\infty dz \  e^{-z \eta} \Psi^{\dagger}(y+z) \Psi^\dagger(x-z)
\end{align*}
The consistency of the algebra requires that the  following equation holds
\begin{align*}
\rl{\Psi(x)\Psi(y)-\Psi(y)\Psi(x)-2 \eta \int_0^\infty dz e^{-z \eta} \Psi(x-z)\Psi(y+z))}\begin{pmatrix}
\text{const.}\\\psi(x_1)\\ \psi(x_1,x_2) \\ \psi(x_1,x_2,x_3) \\ .......\\
\end{pmatrix}
=0
\end{align*}
ie,
\begin{align*}
\begin{pmatrix}
\sqrt{2.1}\rl{\psi(y,x)-\psi(x,y)-2 \eta  \int_0^\infty dz e^{-z \eta}\psi(y+z,x-z)  }\\\sqrt{3.2} \rl{\psi(y,x,x_1)-\psi(x,y,x_1)-2 \eta  \int_0^\infty dz e^{-z \eta}\psi(y+z,x-z,x_1)  }\\ \sqrt{4.3} \rl{\psi(y,x,x_1,x_2)-\psi(x,y,x_1,x_2)-2 \eta  \int_0^\infty dz e^{-z \eta}\psi(y+z,x-z,x_1,x_2)  }\\ .......\\
\end{pmatrix} = 0
\end{align*}
where $ \psi(x_1,x_2,..,x_N) $ is the $N$-anyon wave function. Let the wave function be
\begin{align*}
\psi(x_1,..,x_N)  = \sum_{P\in S_N} \alpha\rl{k_{P(1)},..,k_{P(N)}} e^{i\rl{k_{P(1)}x_1  +...+k_{P(N)} x_N}}
\end{align*} 
where the coefficients satisfy
\begin{align*}
\alpha\rl{...k_j,..k_l,..} = \rl{\frac{k_j-k_l-i\eta}{k_j-k_l+i\eta}} \alpha\rl{...k_l,..k_i,..}
\end{align*}
We have to calculate
\begin{align*}
\rl{\psi(y,x,x_3,..,x_N)-\psi(x,y,x_3,..,x_N)-2 \eta  \int_0^\infty dz e^{-z \eta}\psi(y+z,x-z,x_3,..,x_N)  }
\end{align*}
Substituting the expression for the wave function
\begin{align*}
\sum_{P\in S_N}  &\bigg( \alpha\rl{k_{P(1)},..,k_{P(N)}} e^{i\rl{k_{P(1)}y +k_{P(2)}x+k_{P(3)}x_3 +...+k_{P(N)} x_N}} \\ &-\alpha\rl{k_{P(1)},..,k_{P(N)}} e^{i\rl{k_{P(1)}x +k_{P(2)}y +k_{P(3)}x_3+...+k_{P(N)} x_N}} \\ &-2 \eta  \int_0^\infty dz e^{-z \eta}\alpha\rl{k_{P(1)},..,k_{P(N)}} e^{i\rl{k_{P(1)}\rl{y+z} +k_{P(2)}\rl{x-z}+k_{P(3)}x_3 +...+k_{P(N)} x_N}}  \bigg)\\
& = \sum_{P\in S_N}  \bigg( \alpha\rl{k_{P(1)},..,k_{P(N)}} e^{i\rl{k_{P(1)}y +k_{P(2)}x+k_{P(3)}x_3 +...+k_{P(N)} x_N}} \\ &-\alpha\rl{k_{P(1)},..,k_{P(N)}} e^{i\rl{k_{P(1)}x +k_{P(2)}y +k_{P(3)}x_3+...+k_{P(N)} x_N}} \\ &-\frac{2 i \eta}{k_{P(1)}-k_{P(2)}+i \eta}  \alpha\rl{k_{P(1)},..,k_{P(N)}} e^{i\rl{k_{P(1)}y +k_{P(2)}x+k_{P(3)}x_3 +...+k_{P(N)} x_N}}  \bigg)
\end{align*}
we find that the coefficient of the term $ e^{i\rl{k_{P(1)}x +k_{P(2)}y+k_{P(3)}x_3 +...+k_{P(N)} x_N}}  $ is
\begin{align*}
\alpha\rl{k_{P(2)},k_{P(1)},..,k_{P(N)}} - \rl{\frac{k_{P(1)}-k_{P(2)}-i\eta}{k_{P(1)}-k_{P(2)}+i\eta}} \alpha\rl{k_{P(1)},k_{P(2)},..,k_{P(N)}}
\end{align*}
Using the relation among coefficients, it is easy to see that above term is zero, as expected.
\section{Calculation of the one-particle reduced density matrix}\label{redmat}
The matrix elements of the one-particle reduced density matrix are
\begin{align*}
\rho^{(1)}_{m,n} =  \frac{\sum_{k=0}^{\infty} \bra{0}\Psi_{h_m}  \Psi_{h_k} \ket{\Phi_{j,i}}\bra{\Phi_{j,i}} \Psi_{h_k}^\dagger \Psi^\dagger_{h_n} \ket{0} }{\bra{\Phi_{j,i}}\hat{\mathbf n} \ket{\Phi_{j,i}}}.
\end{align*}
Using the definition of the state $ \ket{\Phi_{j,i}} $, it is rewritten as,
\begin{align*}
\rho^{(1)}_{mn} = \frac{\sum_k \bra{0} \Psi_{h_m} \Psi_{h_k} \Psi_{h_j}^\dagger \Psi_{h_i}^\dagger \ket{0} \bra{0} \Psi_{h_i}\Psi_{h_j}\Psi_{h_k}^\dagger \Psi^\dagger_{h_n} \ket{0}}{2\braket{0|\Psi_{h_i}\Psi_{h_j} \Psi_{h_j}^\dagger \Psi_{h_i}^\dagger |0}}
\end{align*}
To obtain the expression for the one-particle reduced density matrix a generic term of the following form is calculated
\begin{align*}
\bra{0}\Psi_{h_m}\Psi_{h_k} \Psi_{h_j}^\dagger \Psi_{h_i}^\dagger \ket{0} & =\braket{h_k|h_j}\braket{h_m|h_i}+ \braket{h_k|h_i}\braket{h_m|h_j} \\ &-\int_{0}^\infty dz \ 2 \eta e^{-z \eta} \int_{-\infty}^\infty dxdy  h_m^*(y-z) h_k^*(x) h_j(y) h_i(x-z) 
\end{align*}
Using the above formula, the denominator of the one-particle reduced density matrix can be obtained by setting $ m=i $ and $ k=j $. The numerator is calculated below.
\begin{align*}
\sum_k \bra{0} & \Psi_{h_m} \Psi_{h_k} \Psi_{h_j}^\dagger \Psi_{h_i}^\dagger \ket{0} \bra{0} \Psi_{h_i}\Psi_{h_j}\Psi_{h_k}^\dagger \Psi^\dagger_{h_n} \ket{0} \\
&=   \braket{h_m|h_i}\braket{h_n|h_i}+\braket{h_i|h_j} \braket{h_m|h_i}  \braket{h_n|h_j} \\& + \braket{h_j|h_i} \braket{h_m|h_j} \braket{h_i|h_n}+ \braket{h_m|h_j}\braket{h_j|h_n} \\&  -  2 \eta \braket{h_m|h_i}\int_{0}^\infty dz  \int_{-\infty}^\infty dxdy  \ e^{-z \eta}  h_n(y-z) h_j(x) h_j^*(y)h_i^*(x-z)\\
&  - 2 \eta \braket{h_m|h_j} \int_{0}^\infty dz\int_{-\infty}^\infty dxdy \ e^{-z \eta}  h_n(y-z) h_i(x) h_j^*(y) h_i^*(x-z) \\
&- 2 \eta \braket{h_i|h_n}\int_{0}^\infty dz\int_{-\infty}^\infty dxdy  \ e^{-z \eta}  h_m^*(y-z) h_j^*(x) h_j(y) h_i(x-z) \\
&- 2 \eta \braket{h_j|h_n}\int_{0}^\infty dz \int_{-\infty}^\infty dxdy  \ e^{-z \eta}  h_m^*(y-z) h_i^*(x) h_j(y) h_i(x-z)  \\
&+ 4 \eta^2 \int_{0}^\infty dz  dz'  \int_{-\infty}^\infty dx dy dy' \bigg( e^{-(z+z') \eta}  h_m^*(y-z)  h_j(y) h_i(x-z)  \\& \quad \quad \quad \quad \quad \quad \quad \quad  \times h_n(y'-z') h_j^*(y') h_i^*(x-z')  \bigg)
\end{align*}

To calculate the one-particle reduced density matrix, we use the following integrals \cite{gradshteyn2014table}. 
\begin{align*}
 \int_{-\infty}^{\infty} dz & \ e^{-\frac{z^2}{2}-\frac{1}{2} (z-\zeta )^2} H_n(z) H_p(z-\zeta )\\ =&  \frac{1}{{\Gamma (n+1)}}\sqrt{\pi } e^{-\frac{\zeta ^2}{4}} \sqrt{2^n n!} \sqrt{2^p p!} (-\zeta )^{p-n}
 \\& \times	\sqrt{2^{n-p} \Gamma (n+1) \Gamma (p+1)} \, _1\tilde{F}_1\left(-n;-n+p+1;\frac{\zeta
 		^2}{2}\right) \quad , n,p \in \mathbb{N} 
\end{align*}
\begin{align*}
\int_{0}^{\infty} x^{\nu-1} e^{-\beta x^2 -\gamma x}  = \rl{2 \beta}^{-\frac{\nu}{2}} \Gamma\rl{\nu} e^{\frac{\gamma^2}{8\beta}} D_{-\nu} \rl{\frac{\gamma}{\sqrt{2 \beta}}} \ \ , \ \nu >-1
\end{align*}
Here $ \, _1\tilde{F}_1\left(a;b;z\right) $ denotes the regularized confluent hypergeometric function and $  D_{-\nu}(z) $denotes the parabolic cylinder function.
The matrix elements of the one-particle reduced density matrix obtained from the initial state $ \ket{\Phi_{0,0} }$ 
are given below.
\begin{align*}
\rl{ \rho^{(1)}_{0,0}}_{m,n}  =&  \frac{1}{d_1} \bigg(4 \delta_{m0} \delta_{n0} - 4 \eta  \delta_{m0}(-1)^n  \frac{1}{\sqrt{2^n n!}}   \mathfrak{D}(n+1,\eta)
\\&- 4 \eta  \delta_{n0} (-1)^m  \frac{1}{\sqrt{2^m m!}}   \mathfrak{D}(m+1,\eta)\\
& \quad +(-1)^{m+n}  \frac{4 \eta^2 }{\sqrt{2^{m+n} m! n!}}  \sum_{l=0}^{\infty}\bigg(\frac{ 1}{2^l l!}    \mathfrak{D}(m+l+1,\eta)\mathfrak{D}(n+l+1,\eta) \bigg)
\bigg)
\end{align*}
where $ \mathfrak{D}(\nu,x) = \Gamma(\nu) e^{\frac{\eta^2}{4}} D_{-\nu}(x) $ and 
\begin{align*}
d_1 = 4(1-  \eta  \mathfrak{D}(-1,\eta) )
\end{align*} 
The matrix elements of the one-particle reduced density matrix obtained from the initial state $ \ket{\Phi_{1,0} }$ are given below 
\begin{align*}
\rl{ \rho^{(1)}_{1,0}}_{m,n} = & \frac{1}{d_2} \bigg( \delta_{m1} \delta_{n1}+ \delta_{m0}\delta_{n0}  -  \delta_{m1} \frac{\rl{-1}^{n+1} \sqrt{2} \eta }{\sqrt{2^n n!}}   \mathfrak{D}(n+2,\eta)\\
&  - \delta_{m0} \frac{\eta (-1)^n}{\sqrt{2^n n!}}  \rl{ 2    \mathfrak{D}(n+1,\eta)-   \mathfrak{D}(n+3,\eta)} \\
&- \delta_{n1}\frac{\rl{-1}^{m+1} \sqrt{2} \eta }{\sqrt{2^m m!}}    \mathfrak{D}(m+2,\eta)\\
&- \delta_{n0} \frac{\eta (-1)^m}{\sqrt{2^m m!}}  \rl{ 2    \mathfrak{D}(m+1,\eta)-  \mathfrak{D}(m+3,\eta) }  \\
&+ \frac{4 \eta^2 (-1)^{m+n} }{2^{m+n} m! n!}   \sum_{l=0}^{\infty} \frac{1}{2^l l!}\bigg( 2  \mathfrak{D}(n+l+1,\eta) \mathfrak{D}(m+l+1,\eta)\\
&\quad -   \mathfrak{D}(n+l+1,\eta) \mathfrak{D}(m+l+3,\eta) +  2 \mathfrak{D}(n+l+2,\eta) \mathfrak{D}(m+l+2\eta)\\
&\quad -   \mathfrak{D}(n+l+3,\eta) \mathfrak{D}(m+l+1,\eta)\bigg) \bigg)
\end{align*}

where 
\begin{align*}
d_2=&2\rl{1 + \frac{\eta}{2}   \mathfrak{D}(3,\eta)}\\
\end{align*}
\bibliographystyle{unsrt}
\bibliography{paperv2} 
\end{document}